\documentclass[sigconf]{acmart}
\usepackage{booktabs} 
\usepackage{adjustbox}
\usepackage{multirow}
\usepackage{multicol}
\usepackage{bm}
\usepackage{pgfplots}
\usepackage{url}
\usepackage{hyperref}
\usepackage{subfigure}
\usepackage{tablefootnote}

\usepackage{CJKutf8}
\usepackage{amsmath}
\usepackage{amsfonts}
\usepackage{bm}
\usepackage{scalerel}

\pgfplotsset{width=10cm,compat=1.9}

\makeatletter
\def\BState{\State\hskip-\ALG@thistlm}
\makeatother
\usepackage{tkz-euclide}
\usetikzlibrary{shapes.arrows, fadings, automata,tikzmark,decorations.pathreplacing,patterns}
\usetikzlibrary{intersections}
\usetikzlibrary{arrows.meta}
\usetikzlibrary{positioning, fit, mindmap, trees, calc,tikzmark,shapes}
\AtBeginDocument{%
  \providecommand\BibTeX{{%
    \normalfont B\kern-0.5em{\scshape i\kern-0.25em b}\kern-0.8em\TeX}}}

\copyrightyear{2022}
\acmYear{2022}
\setcopyright{acmlicensed}\acmConference[KDD '22]{Proceedings of the 28th ACM SIGKDD Conference on Knowledge Discovery and Data Mining}{August 14--18, 2022}{Washington, DC, USA}
\acmBooktitle{Proceedings of the 28th ACM SIGKDD Conference on Knowledge Discovery and Data Mining (KDD '22), August 14--18, 2022, Washington, DC, USA}
\acmPrice{15.00}
\acmDOI{10.1145/3534678.3539092}
\acmISBN{978-1-4503-9385-0/22/08}
\acmPrice{15.00}
\acmISBN{978-1-4503-XXXX-X/18/06}

\acmSubmissionID{1166}

\DeclareMathOperator{\bE}{\mathbb{E}}
\DeclareMathOperator{\bP}{\mathbb{P}}
\DeclareMathOperator{\cD}{\mathcal{D}}
\usepackage[ruled,vlined]{algorithm2e}
\usepackage{xcolor}

\newcommand{\watchtime}{WatchTime}

\definecolor{tea_green}{RGB}{214, 234, 193}
\definecolor{hint_green}{RGB}{226,246,209}
\definecolor{Madang}{RGB}{190,235,159}
\definecolor{yellow_green}{RGB}{198,222,119}
\definecolor{link_water}{RGB}{221, 232, 250}
\definecolor{celestial_blue}{RGB}{52, 152, 219}
\definecolor{shakespeare}{RGB}{85, 154, 193}
\definecolor{buttermilk}{RGB}{255,242,174}
\definecolor{chardonnay}{RGB}{250,196,114}
\definecolor{rajah}{RGB}{253,180,98}
\definecolor{fog}{RGB}{213, 193, 234}
\definecolor{melon}{RGB}{254,191,181}
\definecolor{sundown}{RGB}{249, 180, 181}
\definecolor{mona_lisa}{RGB}{246,152,134}
\definecolor{salmon}{RGB}{242,131,107}

\definecolor{saltpan}{RGB}{238, 243, 232}
\definecolor{aqua_spring}{RGB}{232, 243, 232}
\definecolor{tea_green}{RGB}{214, 234, 193}
\definecolor{Madang}{RGB}{190,235,159}
\definecolor{fringy_flower}{RGB}{194, 234, 193}
\definecolor{aero_blue}{RGB}{193, 234, 213}
\definecolor{pixie_green}{RGB}{183,214,170}
\definecolor{french_pass}{RGB}{195,232,246}
\definecolor{ice_cold}{RGB}{169,232,220}
\definecolor{pale_turquoise}{RGB}{172,240,242}
\definecolor{cruise}{RGB}{179,226,205}
\definecolor{sail}{RGB}{163,205,235}
\definecolor{spindle}{RGB}{179,205,227}
\definecolor{link_water}{RGB}{221, 232, 250}
\definecolor{periwinkle}{RGB}{203,213,232}
\definecolor{zanah}{RGB}{220, 233, 213}
\definecolor{frostee}{RGB}{217, 231, 214}
\definecolor{opal}{RGB}{199, 221, 211}
\definecolor{jet_stream}{RGB}{188, 214, 210}
\definecolor{skeptic}{RGB}{153, 187, 167}
\definecolor{hint_green}{RGB}{226,246,209}
\definecolor{snow_flurry}{RGB}{230,245,201}
\definecolor{surf_crest}{RGB}{205,230,208}
\definecolor{yellow_green}{RGB}{198,222,119}
\definecolor{cream}{RGB}{255,255,204}
\definecolor{pale_prim}{RGB}{255,255,179}
\definecolor{spring_sun}{RGB}{242,243,195}
\definecolor{portafino}{RGB}{245,237,160}
\definecolor{buttermilk}{RGB}{255,242,174}
\definecolor{cream_brulee}{RGB}{255, 229, 151}
\definecolor{dairy_cream}{RGB}{254,226,189}
\definecolor{champagne}{RGB}{254,217,166}
\definecolor{chardonnay}{RGB}{250,196,114}
\definecolor{manhattan}{RGB}{226,180,125}
\definecolor{rajah}{RGB}{253,180,98}
\definecolor{early_dawn}{RGB}{252,243,218}
\definecolor{egg_shell}{RGB}{238, 234, 215}
\definecolor{selago}{RGB}{243, 232, 243}
\definecolor{quartz}{RGB}{219,223,238}
\definecolor{fog}{RGB}{213, 193, 234}
\definecolor{languid_lavender}{RGB}{222,203,228}
\definecolor{watusi}{RGB}{254,221,207}
\definecolor{coral_andy}{RGB}{243,204,205}
\definecolor{cosmos}{RGB}{248,209,210}
\definecolor{melon}{RGB}{254,191,181}
\definecolor{azalea}{RGB}{234, 193, 194}
\definecolor{beauty_bush}{RGB}{235, 185, 179}
\definecolor{sundown}{RGB}{249, 180, 181}
\definecolor{mona_lisa}{RGB}{246,152,134}
\definecolor{salmon}{RGB}{242,131,107}

\definecolor{summer_sky}{RGB}{58, 151, 233}
\definecolor{chateau_green}{RGB}{72, 179, 96}
\definecolor{matisse}{RGB}{25, 104, 167}
\definecolor{allports}{RGB}{31, 106, 125}
\definecolor{sun_shade}{RGB}{255, 144, 68}
\definecolor{flamingo}{RGB}{237, 88, 85}
\definecolor{studio}{RGB}{128, 91, 160}

\definecolor{maya_blue}{RGB}{102, 204, 255}
\definecolor{feijoa}{RGB}{178,223,138}
\definecolor{sushi}{RGB}{117, 168, 47}
\definecolor{norway}{RGB}{158, 194, 132}
\definecolor{japanese_laurel}{RGB}{53, 116, 40}
\definecolor{see_green}{RGB}{161,228,195}
\definecolor{monte_carlo}{RGB}{135,204,194}
\definecolor{granny_smith_apple}{RGB}{150,214,150}
\definecolor{moss_green}{RGB}{170,216,176}
\definecolor{chateau_green}{RGB}{72, 179, 96}
\definecolor{opal}{RGB}{164,207,190}
\definecolor{acapulco}{RGB}{117, 170, 148}
\definecolor{viridian}{RGB}{55, 137, 122}
\definecolor{amazon}{RGB}{56, 123, 84}
\definecolor{asparagus}{RGB}{123, 160, 91}
\definecolor{fruit_salad}{RGB}{91, 160, 94}
\definecolor{puerto_rico}{RGB}{72, 179, 150}
\definecolor{mountain_meadow}{RGB}{0, 163, 136}
\definecolor{matisse}{RGB}{25, 104, 167}
\definecolor{allports}{RGB}{31, 106, 125}
\definecolor{astral}{RGB}{55, 111, 137}
\definecolor{spring_leaves}{RGB}{46, 83, 117}
\definecolor{biscay}{RGB}{44, 62, 80}
\definecolor{midnight}{RGB}{0, 29, 50}
\definecolor{amethyst}{RGB}{153, 102, 204}
\definecolor{studio}{RGB}{128, 91, 160}
\definecolor{tapestry}{RGB}{194, 109, 132}
\definecolor{atomic_tangerine}{RGB}{255, 153, 102}
\definecolor{amber}{RGB}{255, 191, 0}
\definecolor{casablanca}{RGB}{244, 178, 84}
\definecolor{california}{RGB}{233, 140, 58}
\definecolor{tomato}{RGB}{255, 97, 56} 
\definecolor{alizarin}{RGB}{233, 58, 64}

\definecolor{linen}{RGB}{251, 239, 227}
\definecolor{double_pearl_lusta}{RGB}{253, 242, 208}
\definecolor{oasis}{RGB}{253, 242, 208}
\definecolor{milan}{RGB}{255, 254, 169}
\definecolor{texas}{RGB}{245, 232, 123}
\definecolor{maize}{RGB}{249, 212, 156}

\definecolor{turmeric}{RGB}{211, 178, 76}
\definecolor{saffron}{RGB}{249,193,62}
\definecolor{my_sin}{RGB}{255, 176, 59}
\definecolor{tree_poppy}{RGB}{246, 154, 27}
\definecolor{jaffa}{RGB}{240, 131, 58}
\definecolor{crusta}{RGB}{254, 127, 44}
\definecolor{tahiti_gold}{RGB}{223, 102, 36}
\definecolor{outrageous_orange}{RGB}{255, 100, 45}
\definecolor{safety_orange}{RGB}{254, 106, 0}

\definecolor{azalea}{RGB}{251, 196, 196}
\definecolor{oyster_pink}{RGB}{238,206,205} 
\definecolor{coral_candy}{RGB}{242,208,205} 
\definecolor{baby_pink}{RGB}{246, 194, 192}
\definecolor{petite_orchid}{RGB}{223, 157, 155}
\definecolor{apricot}{RGB}{241,140,122}
\definecolor{NY_pink}{RGB}{228,136,113}
\definecolor{carmine_pink}{RGB}{231, 76, 60}
\definecolor{deep_carmine_pink}{RGB}{236, 50, 67}

\definecolor{wewak}{RGB}{244, 143, 150}
\definecolor{light_coral}{RGB}{244, 127, 123}
\definecolor{bittersweet}{RGB}{255,111,105}
\definecolor{carnation}{RGB}{245, 80, 86}
\definecolor{flamingo}{RGB}{237, 88, 85}
\definecolor{sunset_orange}{RGB}{242,89,75}
\definecolor{ku_crimson}{RGB}{243, 0, 25}
\definecolor{amaranth}{RGB}{234,46,73}
\definecolor{valencia}{RGB}{214, 87, 70}
\definecolor{chilean_fire}{RGB}{215, 87, 44}
\definecolor{mexican_red}{RGB}{170, 41, 37}

\definecolor{napa}{RGB}{163, 154, 137}

\definecolor{athens_gray}{RGB}{236, 240, 241}
\definecolor{gallery}{RGB}{240,240,240}
\definecolor{mercury}{RGB}{230,230,230}
\definecolor{platinum}{RGB}{228,228,228}
\definecolor{silver}{RGB}{191,191,191}
\definecolor{aluminum}{RGB}{153,153,153}
\definecolor{ship_gray}{RGB}{77,77,77}
\definecolor{tuatara}{RGB}{67, 67, 67}

\definecolor{malibu}{RGB}{110, 180, 240}
\definecolor{celestial_blue}{RGB}{52, 152, 219}
\definecolor{curious_blue}{RGB}{41, 128, 185}
\definecolor{french_blue}{RGB}{0, 112, 182}
\definecolor{matisse}{RGB}{25, 104, 167}
\definecolor{shakespeare}{RGB}{85, 154, 193}
\definecolor{seagull}{RGB}{128,177,211}
\definecolor{jelly_bean}{RGB}{45, 126, 150}
\definecolor{venice_blue}{RGB}{87, 135, 105}
\definecolor{boston_blue}{RGB}{68, 147, 161}

\definecolor{turquoise}{RGB}{41,217,194}
\definecolor{java}{RGB}{2,190,196}
\definecolor{riptide}{RGB}{141,211,199}
\definecolor{mountain_meadow}{RGB}{0, 163, 136}
\definecolor{free_speech_aquamarine}{RGB}{0, 156, 114}

\definecolor{cosmic_latte}{RGB}{222, 247, 229}
\definecolor{chinook}{RGB}{163, 232, 178}
\definecolor{padua}{RGB}{121, 189, 143}
\definecolor{ocean_green}{RGB}{79, 176, 112}
\definecolor{pastel_green}{RGB}{107, 227, 135}
\definecolor{chateau_green}{RGB}{69, 191, 85}
\definecolor{RoyalBlue}{RGB}{69, 191, 85}
\definecolor{pigment_green}{RGB}{0, 175, 79}
\definecolor{fern}{RGB}{101,197,117}
\definecolor{killarney}{RGB}{56, 113, 66}

\definecolor{quartz}{RGB}{219,223,238}
\definecolor{spring_sun}{RGB}{242,243,195}
\definecolor{dairy_cream}{RGB}{254,226,189}
\definecolor{surf_crest}{RGB}{205,230,208}
\definecolor{french_pass}{RGB}{195,232,246}
\definecolor{cosmos}{RGB}{248,209,210}
\definecolor{portafino}{RGB}{245,237,160}
\definecolor{sail}{RGB}{163,205,235}
\definecolor{hint_green}{RGB}{226,246,209}
\definecolor{bittersweet}{RGB}{255,111,105}
\definecolor{java}{RGB}{2,190,196}
\definecolor{ice_cold}{RGB}{169,232,220}
\definecolor{bgc}{RGB}{245,245,245}
\definecolor{tuatara}{RGB}{67, 67, 67}
\definecolor{aluminum}{RGB}{153,153,153}
\definecolor{silver}{RGB}{191,191,191}
\definecolor{platinum}{RGB}{228,228,228}
\definecolor{mercury}{RGB}{230,230,230}
\definecolor{gallery}{RGB}{240,240,240}
\definecolor{free_speech_aquamarine}{RGB}{0, 156, 114}
\definecolor{sun_shade}{RGB}{255, 144, 68}
\definecolor{fern}{RGB}{101,197,117}
\definecolor{french_blue}{RGB}{0, 112, 182}
\definecolor{matisse}{RGB}{25, 104, 167}
\definecolor{sushi}{RGB}{117, 168, 47}
\definecolor{shakespeare}{RGB}{85, 154, 193}
\definecolor{egg_shell}{RGB}{238, 234, 215}
\definecolor{carnation}{RGB}{245, 80, 86}
\definecolor{flamingo}{RGB}{237, 88, 85}
\definecolor{jet_stream}{RGB}{188, 214, 210}
\definecolor{jelly_bean}{RGB}{45, 126, 150}
\definecolor{tree_poppy}{RGB}{246, 154, 27}
\definecolor{deep_carmine_pink}{RGB}{236, 50, 67}
\definecolor{copper_rust}{RGB}{155, 64, 74}
\definecolor{midnight}{RGB}{0, 29, 50}
\definecolor{chilean_fire}{RGB}{215, 87, 44}
\definecolor{puerto_rico}{RGB}{94, 194, 166}
\definecolor{japanese_laurel}{RGB}{53, 116, 40}
\definecolor{fire_engine_red}{RGB}{206, 37, 51}
\definecolor{ku_crimson}{RGB}{243, 0, 25}
\definecolor{turmeric}{RGB}{211, 178, 76}
\definecolor{tahiti_gold}{RGB}{223, 102, 36}
\definecolor{outrageous_orange}{RGB}{255, 100, 45}
\definecolor{crusta}{RGB}{254, 127, 44}
\definecolor{safety_orange}{RGB}{254, 106, 0}
\definecolor{pigment_green}{RGB}{0, 175, 79}
\definecolor{jaffa}{RGB}{240, 131, 58}
\definecolor{jet_stream}{rgb}{0.69,0.61,0.85}
\definecolor{jelly_bean}{rgb}{0.47,0.32,0.66}
\definecolor{azalea}{RGB}{251, 196, 196}
\definecolor{sundown}{RGB}{249, 180, 181}
\definecolor{light_coral}{RGB}{244, 127, 123}
\definecolor{wewak}{RGB}{244, 143, 150}

\definecolor{biscay}{RGB}{44, 62, 80}
\definecolor{carmine_pink}{RGB}{231, 76, 60}
\definecolor{athens_gray}{RGB}{236, 240, 241}
\definecolor{celestial_blue}{RGB}{52, 152, 219}
\definecolor{curious_blue}{RGB}{41, 128, 185}
\definecolor{my_sin}{RGB}{255, 176, 59}
\definecolor{viridian}{RGB}{70, 137, 102}
\definecolor{tomato}{RGB}{255, 97, 56}
\definecolor{mountain_meadow}{RGB}{0, 163, 136}
\definecolor{padua}{RGB}{121, 189, 143}
\definecolor{killarney}{RGB}{56, 113, 66}
\definecolor{ocean_green}{RGB}{79, 176, 112}
\definecolor{pastel_green}{RGB}{107, 227, 135}
\definecolor{chinook}{RGB}{163, 232, 178}
\definecolor{cosmic_latte}{RGB}{222, 247, 229}
\definecolor{chateau_green}{RGB}{69, 191, 85}
\definecolor{RoyalBlue}{RGB}{69, 191, 85}

\definecolor{blue0}{RGB}{240,249,232}
\definecolor{blue1}{RGB}{204,235,197}
\definecolor{blue2}{RGB}{168,221,181}
\definecolor{blue3}{RGB}{123,204,196}
\definecolor{blue4}{RGB}{78,179,211}
\definecolor{blue5}{RGB}{43,140,190}
\definecolor{blue6}{RGB}{8,88,158}

\definecolor{yellow0}{RGB}{255,255,212}
\definecolor{yellow1}{RGB}{254,227,145}
\definecolor{yellow2}{RGB}{254,196,79}
\definecolor{yellow3}{RGB}{254,153,41}
\definecolor{yellow4}{RGB}{236,112,20}
\definecolor{yellow5}{RGB}{204,76,2}
\definecolor{yellow6}{RGB}{140,45,4}

\begin{document}

\title{Deconfounding Duration Bias in Watch-time Prediction for Video Recommendation}


\author{Ruohan Zhan}
\authornote{This work was done during employment at Kuaishou Technology.}
\affiliation{%
  \institution{Hong Kong University of Science and Technology}
  \city{Hong Kong SAR}
  \country{China}}
\email{ruohanzhan@gmail.com}

\author{Changhua Pei}
\affiliation{%
  \institution{Kuaishou Technology}
  \city{Beijing}
  \country{China}}
\email{peichanghua@kuaishou.com}

\author{Qiang Su}
\affiliation{%
  \institution{Kuaishou Technology}
  \city{Beijing}
  \country{China}}
\email{suqiang@kuaishou.com}

\author{Jianfeng Wen}
\affiliation{%
  \institution{Kuaishou Technology}
  \city{Beijing}
  \country{China}}
\email{wenjianfeng@kuaishou.com}

\author{Xueliang Wang}
\affiliation{%
  \institution{Kuaishou Technology}
  \city{Beijing}
  \country{China}}
\email{wangxueliang03@kuaishou.com}

\author{Guanyu Mu}
\affiliation{%
  \institution{Kuaishou Technology}
  \city{Beijing}
  \country{China}}
\email{muguanyu@kuaishou.com}

\author{Dong Zheng}
\affiliation{%
  \institution{Kuaishou Technology}
  \city{Beijing}
  \country{China}}
\email{zhengdong@kuaishou.com}

\author{Peng Jiang}
\affiliation{%
  \institution{Kuaishou Technology}
  \city{Beijing}
  \country{China}}
\email{jiangpeng@kuaishou.com}

\renewcommand{\shortauthors}{Zhan and Pei, et al.}


\begin{abstract}
Watch-time prediction remains to be a key factor in reinforcing user engagement via video recommendations. It has become increasingly important given the  ever-growing popularity of online videos. However, prediction of watch time not only depends on the match between the user and the video but is often mislead by the \emph{duration} of the video itself. With the goal of improving watch time,  recommendation is always biased towards videos with long duration. Models trained on this imbalanced data face the risk of  bias amplification, which misguides platforms to over-recommend videos with long duration but overlook the underlying user interests.

  This paper presents the first work to study duration bias in watch-time prediction for video recommendation. We  employ a \emph{causal graph} illuminating that   duration is a \emph{confounding} factor that concurrently affects video exposure and watch-time prediction---the first effect on video causes the bias issue and should be eliminated, while the second effect on watch time originates from video intrinsic  characteristics and  should be preserved. To remove the undesired bias but leverage the natural effect, we propose a \textbf{D}uration-\textbf{D}econfounded \textbf{Q}uantile-based (\textbf{D2Q}) watch-time prediction framework, which allows for scalability to perform on industry production systems. Through extensive offline evaluation and live experiments, we showcase the effectiveness of this duration-deconfounding   framework by  significantly outperforming the state-of-the-art baselines. We have fully launched our approach on Kuaishou App, which has substantially improved real-time video consumption  due to more accurate watch-time predictions.   

\end{abstract}




\maketitle

\section{Introduction}

\begin{figure}[t]
    \centering
    \subfigure[View of one video.]{\includegraphics[width=0.2\textwidth]{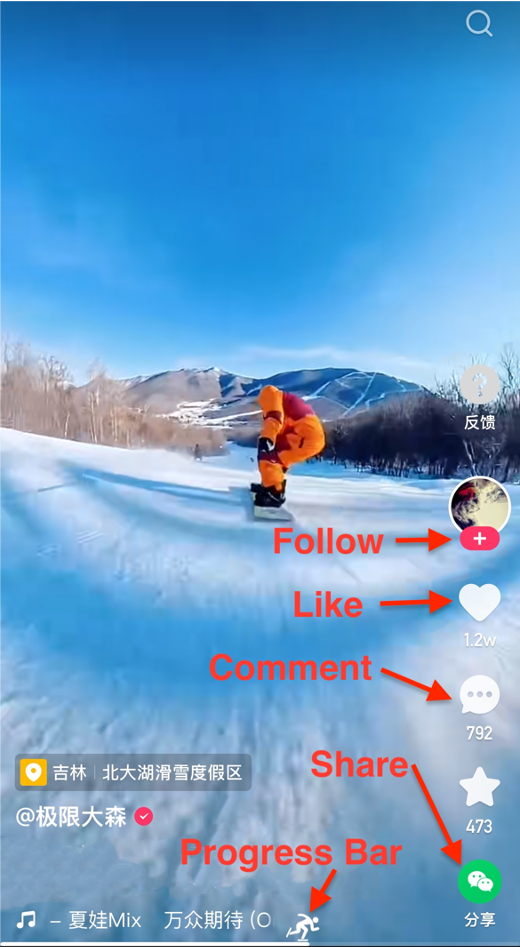}}  
    \subfigure[Scrolling between two videos.]{\includegraphics[width=0.188\textwidth]{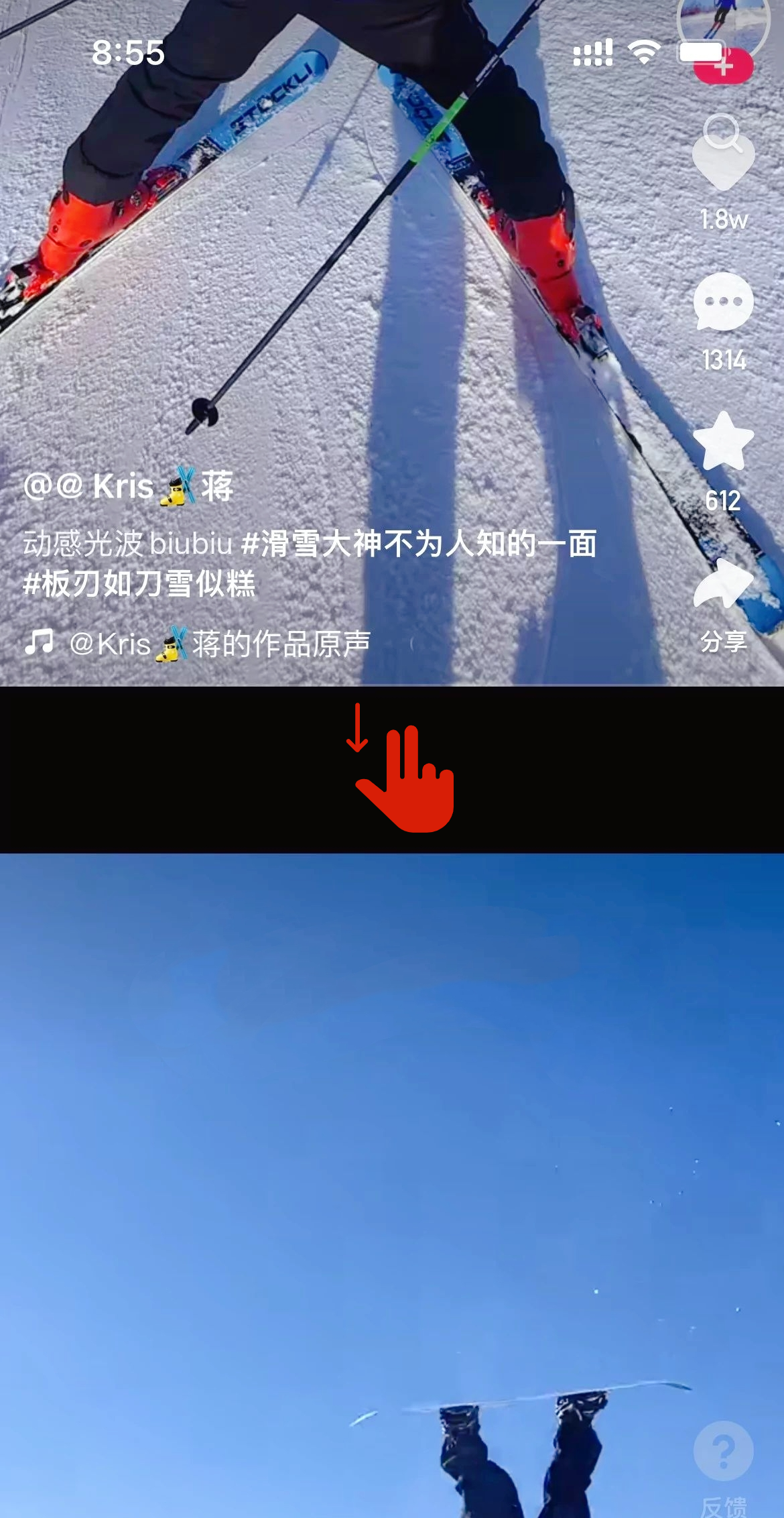}} 
    \caption{A screenshot of the video recommendation module on Kuaishou App. Videos are displayed in the full-screen manner for  immersive user experience.}
    \label{fig:kuaishou-app}
\end{figure}

The rise of online video consumption has drawn growing efforts to optimize recommender systems of internet-based Video on Demand (VOD) systems such as YouTube and streaming players such as  TikTok, Instagram Reels, and Kuaishou (see demonstration in Figure~\ref{fig:kuaishou-app}).  
As such, a main goal is to improve the amount of time users spend on watching the videos, referred to as expected \emph{watch time}~\cite{covington2016deep}. Watch time is a dense signal existing in each video view that concerns every user and video on the platform and represents a scarce resource of user attention that companies compete for. It is thus crucial to accurately estimate watch time on candidate videos when a user arrives. Accurate predictions enable the platform to recommend videos with potentially large watch time to improve user engagement, which directly drives the critical production metric of daily active user (DAU) and thereby the revenue growth.

\begin{figure}
    \centering
    \includegraphics[width=0.943\linewidth]{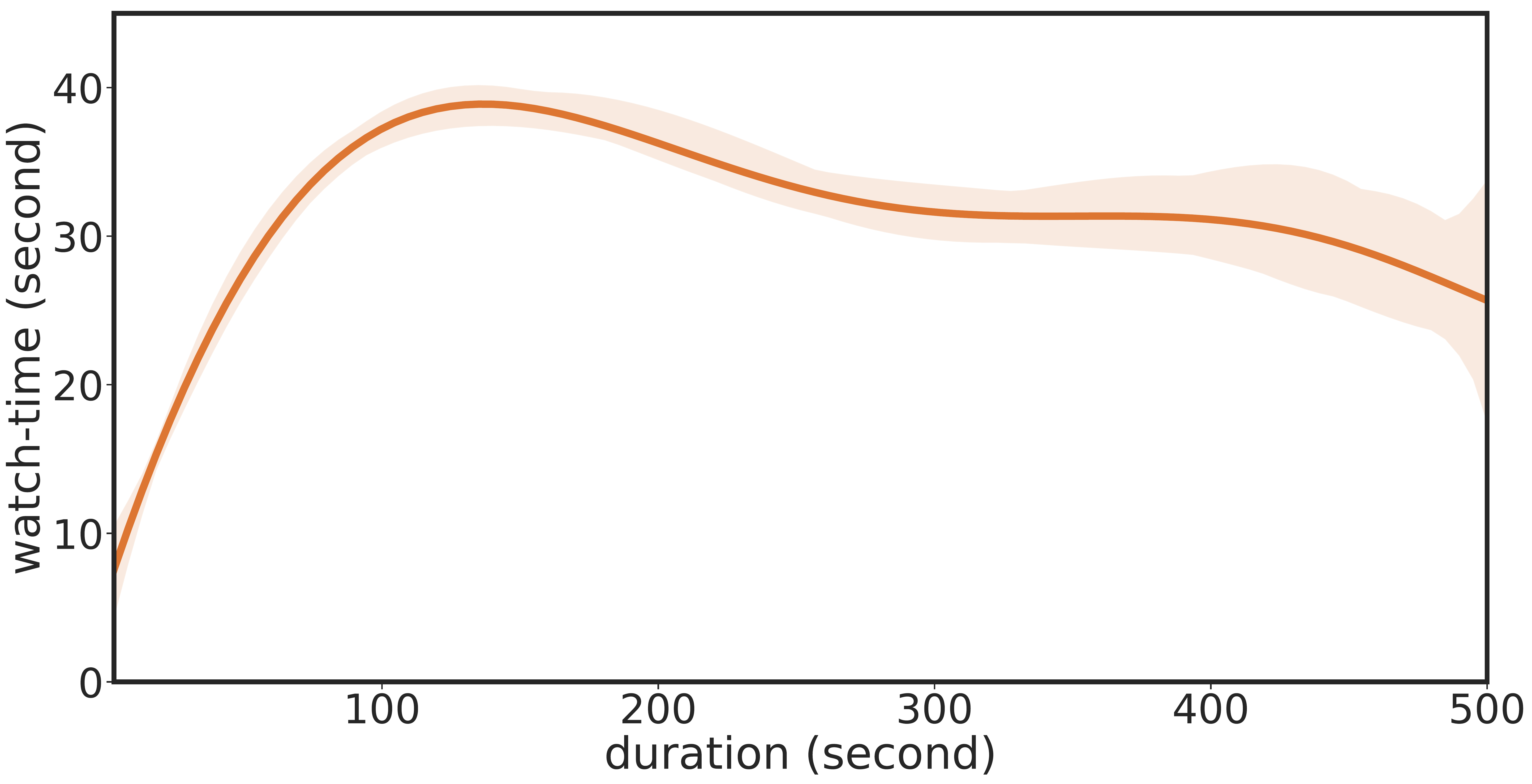}
    \caption{ 60$^{th}$ percentile watch-time on videos with respect to duration. Data is  collected from Kuaishou App with sample size surpassing $20$ billion. The spanned area denotes the $99.99\%$ confidence interval of watch time.}
    \label{fig:duration-watchtime}
\end{figure}

\begin{figure}
    \centering
    \includegraphics[width=0.96\linewidth]{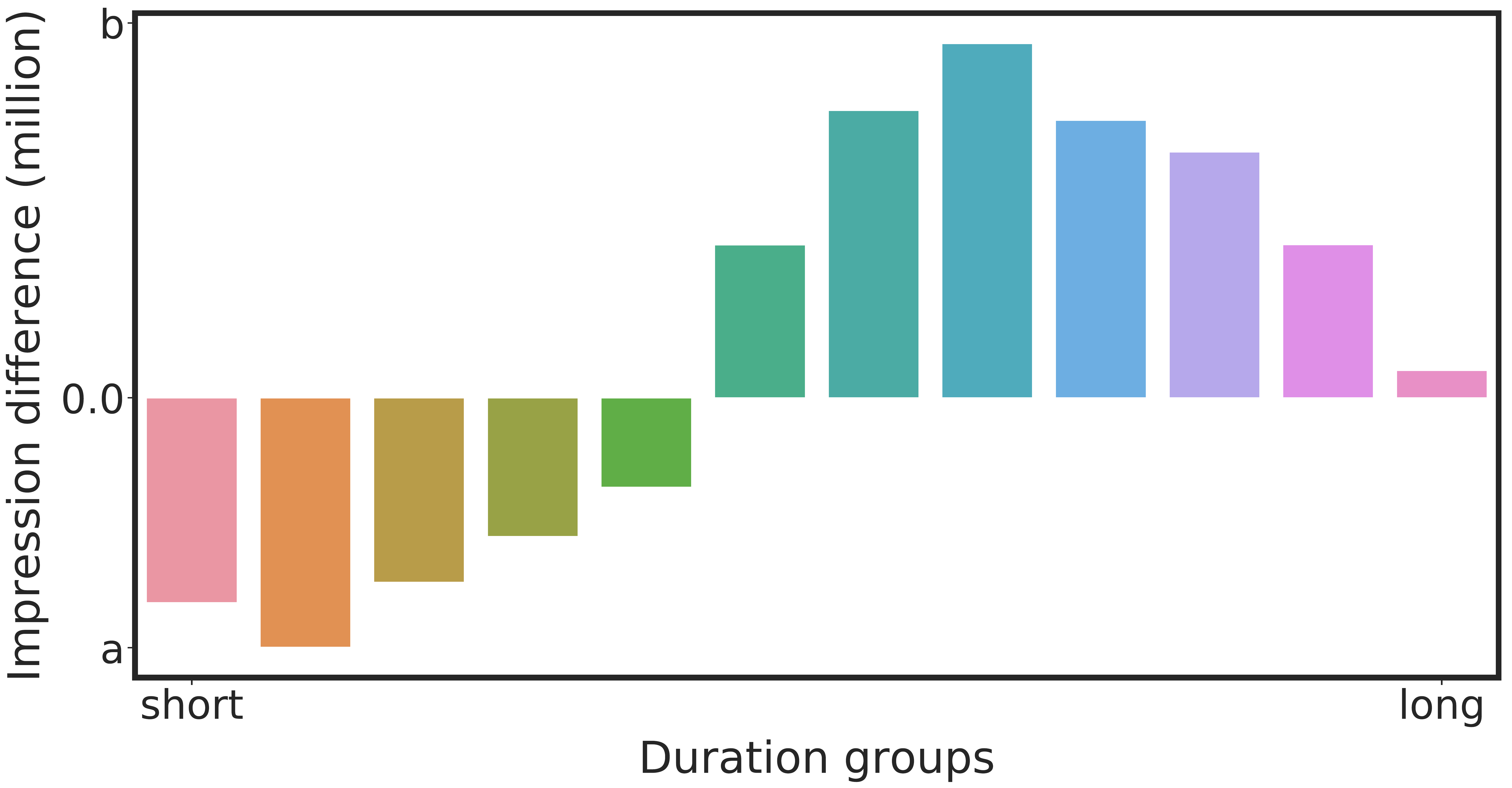}
    \caption{
    Change of video impression over 11 months on Kuaishou App per video duration. Bins are sorted in the ascending order of duration. The height of bin represents the difference of impression counts over this period. The absolute values have been omitted for confidentiality purposes. With the platform's goal to improve watch time, impression is progressively biased towards videos with long duration.}
    \label{fig:duration-impression}
\end{figure}

Watch time is mainly affected by two factors. As known, it is largely governed by how interested the user is in the video and  can be zero when there is no interest match at all \cite{guo2019multi,wang2019overview}. Meanwhile, \emph{duration} of a video itself (\emph{i.e.}, length of the video) also plays a significant role in determining how long the user spends on the video. Figure \ref{fig:duration-watchtime}  shows that user watch-time is positively correlated with video duration. As a result, standard watch-time prediction models often use duration together with other video characteristics as feature inputs to make predictions \cite{covington2016deep,davidson2010youtube}. However, such practice unfortunately results in a bias issue  in many recommender systems. Figure \ref{fig:duration-impression} demonstrates that recommendation is over time progressively based towards videos with long duration, due to the platform's goal of maximizing user watch time. As a result, videos with longer duration are likely to be over-exposed, such that user real interest is undervalued in recommendations. More severely, models trained on such imbalanced data will amplify the duration bias  due to the system's feedback loop \cite{steck2018calibrated}, which undesirably harms the diversity and personalization in ideal recommendation. 

Despite the high prevalence, duration bias is much less explored as compared to many other biases that are caused by item popularity or position  in recommendation studies  \cite{biega2018equity,mehrotra2018towards,beutel2019fairness,zehlike2020reducing,zheng2021disentangling,zhang2021causal,wang2021deconfounded}. With the goal of maximizing user watch-time, recommender systems may learn spurious correlation between duration and watch-time; thus  videos with longer duration are more likely to be shown  even though they may fail to match the user interest well. On the other hand,  videos with long duration usually have larger sample size resulted from existing imbalanced exposure, which may dominate model learning such that model performance varies across duration. 

This paper presents the first work in studying duration bias in watch-time prediction. We employ a direct acyclic graph (named as \emph{causal graph} \cite{pearl2009causality}) to characterize the causal relationship regarding duration in watch-time prediction, modeled by Figure \ref{fig:causal}(a). Specifically, duration---served as a \emph{confounding} factor~\cite{pearl2009causality}---simultaneously affects  both watch-time prediction and video exposure. The first effect of duration on watch-time shows that users tend to spend more time watching videos with intrinsically longer duration, which is a natural effect and should be captured by watch-time prediction models. The second effect from duration to video, however, is a bias term that plagues many watch-time prediction models. Such effect suggests that duration influences the likelihood of video impression, which represents model's unfair preference on videos with long duration and should be eliminated. Such explicit modeling of duration effects, in contrast to previous works that only use duration as features for watch-time prediction,  allows us to remove the undesired bias but preserve the true influence. 

To deal with the duration bias, we follow the principle of \emph{backdoor adjustment} \cite{pearl2012calculus} and intervene the causal graph of watch-time prediction  to remove the undesired effect from duration to video exposure, as characterized in Figure \ref{fig:causal}(b). We note that the effect from duration to watch-time is preserved, since such relationship is intrinsic and should be leveraged in prediction.  Operationally, we split the training data into equal parts with respect to duration; and for each duration group, we learn a  regression model to  predict group-wise watch-time quantiles, where the labels are determined by  the original watch-time values and the empirical cumulative distribution of watch time in the corresponding group. Such quantile prediction enables model parameter sharing across duration groups, bringing in benefits on scalability.  Together, we summarize our contributions as below:
\begin{itemize}
    \item \textbf{Causal Formulation of Duration Bias in Watch-time Prediction.} We employ a  causal graph to formalize the overlooked but widely existing issue of duration bias  in watch-time prediction. We point out that duration is a confounder that affects both watch-time prediction and video exposure, of which the former is intrinsic and should be preserved, while the latter is bias and should be eliminated. 
    \item \textbf{Adjusting for Duration  with Scalability.} Guided by backdoor adjustment, we split data based on duration and fit a watch-time prediction model for each duration group to remove the duration bias on video exposure. We modify the watch-time labels with regard to duration to allow for parameter sharing across groups and gain scalability.  
    \item \textbf{Extensive Offline Evaluation.} We conduct a series of offline evaluation on data collected from Kuaishou App to demonstrate the advantages of our model over existing baselines. We further do ablation studies on the number of duration groups, enlightening that with  group number increasing, our model performance firstly improves (thanks to duration de-biasing) and then declines (due to increased estimation error from reduced groupwise sample size). 
    \item \textbf{Benefits in Live Experiments.} We further implement our approach in live experiments to facilitate video recommendations on Kuaishou  platform, showing that by removing the undesired duration bias, our approach improves watch-time prediction accuracy and  contributes to optimized real-time video consumption as compared to existing strategies.
\end{itemize}

\section{Related Work}
\paragraph{Watch-time Prediction} 
It is crucial for  many industry recommender systems to accurately predict watch time, which is one of the most representative metrics of user engagement \cite{tang2017popularity,covington2016deep,wu2018beyond}. However, in contrast to other metrics such as Click-Through-Rate (CTR) \cite{richardson2007predicting,wang2019overview,zhou2018deep,chen2021end,qu2016product,pi2020search}, there are fewer research endeavors focused on this area. 
Remarkably,  \cite{covington2016deep} provides one of the industry standard solutions to watch-time prediction, where the regression problem is transformed into Weighted Logistic Regression (which shall be referred to as WLR for the remaining of this paper), and the impressed  videos are weighted with the actual watch time; in this way, the learned odds are approximately the expected watch time (detailed derivation is shown in Appendix~\ref{appendix:vtr}). However,  videos with long duration---which has  positive correlation with  watch time as shown in Figure \ref{fig:duration-watchtime}---often get larger sample weights during model training, amplifying the duration bias. Beside, such approach cannot be directly adapted to streaming services such as TikTok and Kuaishou that provide full-screen video content for immersive user experience, where there are no nominal unimpressed/unclicked  samples---every video sample has been shown to a user and  is thus impressed. In this paper, we revise the method in \cite{covington2016deep} to adapt to  both streaming services (TikTok and Kuaishou) and  scenarios as YouTube with user behaviors of click and watch. The modified method (\textit{i.e.,} WLR) acts as one of baselines in experiment Section~\ref{sec:offline} and Section~\ref{sec:live} .

\paragraph{Bias in recommender systems} Our work is also closely related to a growing literature focused on addressing biases in recommender systems~\cite{chen2020bias}. One line of such strives to  mitigate systematic biases that stand in opposite to fairness~\cite{wu2021fairrec} and social welfare~\cite{ge2021towards}, via promoting equity of attention for items  or improving model performances for subgroups \cite{biega2018equity,mehrotra2018towards,beutel2019fairness,zehlike2020reducing}. Another line focuses on dealing with algorithmic biases to improve model performance and break bias amplification resulted from feedback loop, which we view our work as complementary to. Methods in this line can be broadly categorized into three classes: (i) \emph{causal embedding}, where researchers decompose item embedding with respect to different causal effects and learn each embedding using a curated bias-free dataset  associated with the corresponding effect \cite{zheng2021disentangling,bonner2018causal,guo2019pal}; (ii) \emph{inverse propensity weighting}, where researchers reweights samples following the rule of importance sampling \cite{horvitz1952generalization,schnabel2016recommendations,christakopoulou2020deconfounding} to correct for the distributional shift in training data, such that the learned model captures the bias-free signals to generalize to unseen distributions \cite{schnabel2016recommendations,wang2018position,gruson2019offline}; and (iii) \emph{causal intervention}, where researchers intervene the causal relationship that causes the bias and
add adjustment to eliminate this harmful effect for more accurate estimation \cite{zhang2021causal,wang2021deconfounded,yang2021top}. Our work is classified into the third category, where with causal intervention, we remove the undesired effect of duration on video but preserve the desired effect of duration on watch time.

\paragraph{Causal Intervention} We finally review a relevant line in causality-related observational studies \cite{porta2008dictionary}. The causal relationships among variables are captured by a directed acyclic graph, named as \emph{causal graph}, where nodes denote variables and directional edges denote causal effects \cite{pearl2009causality}. Bias often arises when a model fails to account for a variable---referred to as \emph{confounder}---which simultaneously affects both the feature variables and the outcome \cite{mickey1989impact,pearl2009causality}. In our case, this variable is the duration that has the confounding effects on both video and watch time. To deal with the bias and deconfound the problematic variable, a standard approach is to conduct \emph{do-calculus} that specifies the value of relevant variables, named as \emph{backdoor adjustment}, such that the intervened causal graph eliminates the edge of undesired causal effect \cite{pearl2012calculus}. Such approach has been widely used to estimate causal effect in various domains across healthcare \cite{gupta2021estimating}, bioinformatics \cite{le2013inferring}, and socioeconomics \cite{wang2020development}. More recently, in recommender systems, it receives growing applications to adjust for the item popularity bias for CTR prediction \cite{zhang2021causal,wang2021deconfounded}.

\section{Causal Model of Watch-time Prediction}
\begin{figure}
    \centering
    \includegraphics[width=0.5\textwidth]{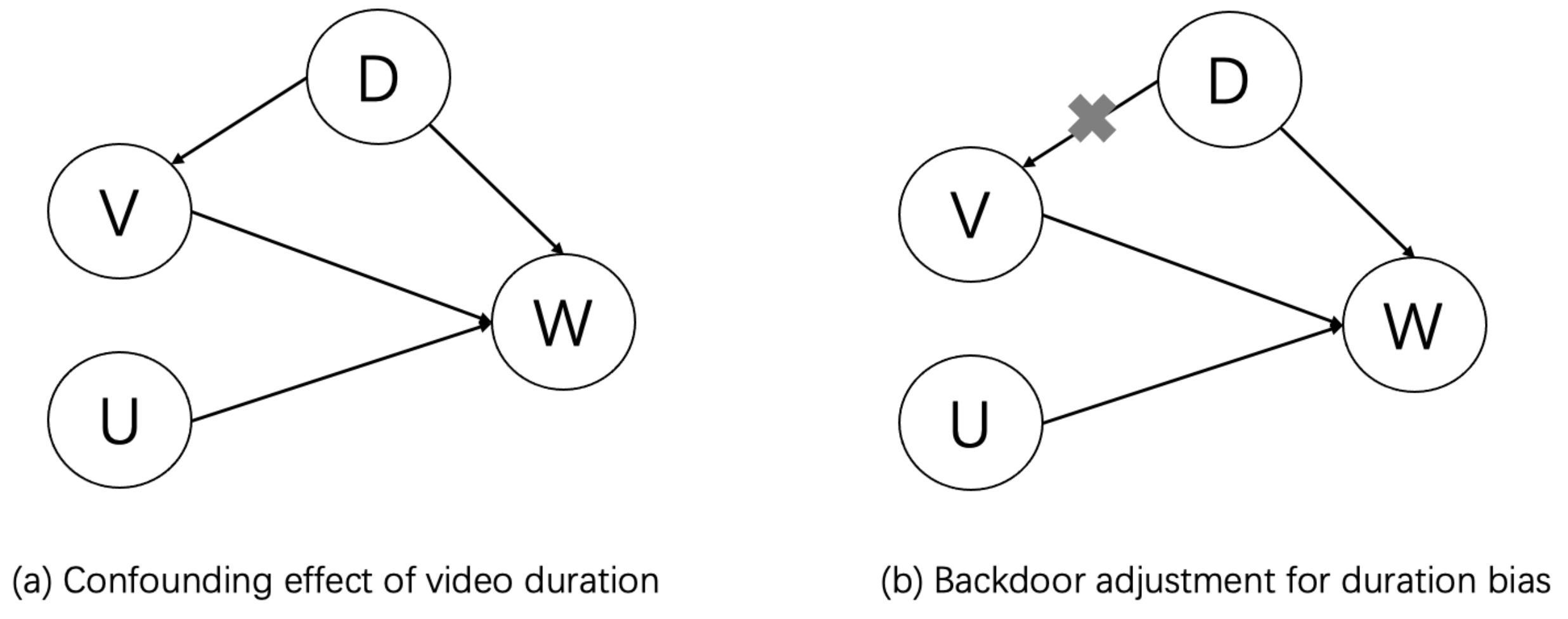}
    \caption{Causal graphs of watch-time prediction: $U$--user, $V$--video, $D$--duration, $W$--watch time. Figure (a) models the confounding effect of video duration on both video exposure and watch-time prediction. Figure (b) uses  backdoor adjustment to deconfound duration and remove its  effect on video.}
    \label{fig:causal}
\end{figure}

Our goal is to predict   watch time of a user when she/he is recommended with a video. 
We start by formulating the problem via a causal graph that characterizes the relationships among user, video, duration, watch-time, and in particular, the confounding effects of duration on both video exposure and watch-time prediction in most recommender systems, as shown in Figure \ref{fig:causal}(a):
\begin{itemize}
    \item $U$ denotes user representation, including user demographics, instantaneous context,  historical interactions, etc.
    \item $V$ denotes video representation, including video topics, etc.
    \item $D$ denotes video duration, i.e., the length of the video.
    \item $W$ denotes the time user spends on watching the video.
    \item $\{U,V\}\rightarrow W$ capture the \emph{interest} effect  on watch-time, which measures how user is interested in the video.
    \item $D\rightarrow W$ captures the \emph{duration} effect on watch time, which suggests that  when two videos match user interest comparably, longer video  may receive longer watch time.
    \item $D\rightarrow V$ implies that duration will affect video exposure. Recommender systems often have unfair preference on videos with long duration, due to the goal of maximizing user watch time; such bias is amplified by feedback loop,  as shown by Figure \ref{fig:duration-impression}. Besides, duration may affect mode training since (i) sample size varies with duration---videos with long duration usually have larger sample size, on which the prediction model  has better performance; and (ii) videos with different duration may receive different sample weights in standard  models such as WLR \cite{covington2016deep}, which affects the allocation of gradient in model training. 
\end{itemize}
Notably, 
the causal graph in Figure \ref{fig:causal}(a) demonstrates that
duration is a confounder that affects watch-time via two paths: $D\rightarrow W$ and $D\rightarrow V \rightarrow W$. The first path suggests that duration has a direct causal relationship with watch time, which should be captured by watch-time prediction models since users tend to spend more time on watching long videos versus short ones. However, the second path implies that video exposure is undesirably affected by its duration, and thus the video distribution is biased towards long videos; and if not eliminated, predictions would face the risk of bias amplification by the feedback loop of recommender systems.


\section{Backdoor Adjustment for Duration Bias}

In this section, we follow the principle of backdoor adjustment to deconfound duration, where we remove the bias from duration on video  but preserve the effect from duration on watch time. We propose a scalable watch-time prediction framework that is Duration-Deconfounded and Quantile-based (D2Q), manifested by 
\begin{enumerate}
    \item[i.] Spitting data  based on duration  to remove duration bias;
    \item[ii.] Fitting watch-time quantiles instead of the original values to enable parameter sharing across  groups for scalability. 
\end{enumerate}
 We summarize our training  and inference procedures  in Algorithms \ref{alg:training}  and \ref{alg:inference} respectively.

\begin{algorithm}
\caption{\textbf{Training of D2Q}: Duration-Deconfounded Quantile-based Watch Time Prediction}
\label{alg:training}
\textbf{Input}: training data $\{(u_i,v_i,d_i,w_i)\}_{i=1}^n$.
\begin{enumerate}
    \item Compute empirical quantiles of duration $\{d_i\}_{i=1}^n$ to determine the  duration groups $\{\cD_k\}_{k=1}^{M}$.
    \item Split data $\{(u_i,v_i,d_i,w_i)\}_{i=1}^n$ into $M$ equal parts based on $\{\cD_k\}_{k=1}^{M}$.
    \item For each duration group $\cD_k$, compute the empirical cumulative distribution of watch time $\widehat{\Phi}_k$ on data $\{w_i: d_i\in\cD_k\}$.
    \item Solve the watch-time quantile prediction model $h$ using all samples:
        \begin{equation*}
     h = \arg\min_{h'}  \sum_{\{(u_i, v_i, w_i)\}_{i=1}^n}\big(h'(u_i, v_i) - \widehat{\Phi}_{k_i}(w_i)\big)^2,
        \end{equation*}
        where $k_i$ is the duration group of sample $i$ such that $d_i \in \cD_{k_i}$. 

\end{enumerate}
\textbf{Output}: duration groups $\{\cD_k\}_{k=1}^{M}$,  watch-time quantile prediction model $h$.
\end{algorithm}

\begin{algorithm}
\caption{\textbf{Inference of D2Q}: Duration-Deconfounded Quantile-based Watch Time Prediction}
\label{alg:inference}
\textbf{Input}: user-video pair $(u_0,v_0)$ to be inquired, duration groups $\{\cD_k\}_{k=1}^{M}$,  watch-time quantile prediction model $h$.
\begin{enumerate}
    \item Find the corresponding duration group $\cD_{k_0}$ for video $v_0$.
    \item Estimate watch time $\widehat{w}_0 = \widehat{\Phi}_{k_0}^{-1}\big(h(u_0,v_0)\big)$.
\end{enumerate}
\textbf{Output}: estimated watch-time $\widehat{w}_0$.
\end{algorithm}

\subsection{Deconfounding Duration}
 Following the \emph{do}-calculus, we block the duration effect on video exposure by removing  edge $D\rightarrow V$, as illustrated by the deconfounded causal graph $G_1$ in Figure \ref{fig:causal}(b). We frame the watch-time prediction model as $\bE[W|do(U,V)]$ and have
\begin{equation}
\label{eq:backdoor}
    \begin{split}
        \bE[W|do(U,V)] & = \bE_{G_1}[W|U,V] \\
        & \stackrel{\mbox{(i)}}{=}\sum_{d}\bP_{G_1}(D=d|U,V)\bE_{G_1}[W|U,V, D=d] \\
        & \stackrel{\mbox{(ii)}}{=}  \sum_{d}\bP_{G_1}(D=d)\bE_{G_1}[W|U,V, D=d]\\
         & \stackrel{\mbox{(iii)}}{=}  \sum_{d}\bP(D=d)\bE[W|U,V, D=d],
    \end{split}
\end{equation}
where (i) is by law of total expectation \cite{weiss2006course}; (ii) is because $D$ is independent of $\{U, V\}$ with the intervention that removes edge $D\rightarrow V$ in graph $G_1$; and (iii) is because such intervention does not change the distribution of $W$ conditioning on $\{U,V,D\}$, and the marginal distribution of $D$ remains the same. 

Equation \eqref{eq:backdoor} sheds light on the design to deconfound duration: one can estimate $\bP(D)$ and $\bE[W|U,V,D]$ separately and then combine them together to construct the final estimation. In this paper, we propose to discretize the duration distribution $\bP(D)$  into disjoint groups  and fit group-wise watch-time prediction model $\bE[W|U,V,D]$ to complete the estimation.

\subsection{Data-Splitting based on Duration Quantiles}
We now present  a general framework to estimate the watch-time with duration deconfounded, depicted by Figure \ref{fig:causal}(b). The high-level idea is to split data based on duration and construct group-wise watch-time estimation 
to debiase duration on video exposure. 

Specifically, to block edge $D\rightarrow V$,  we split training samples into $M$ equal parts based on duration quantiles, which discretizes the distribution $\bP(D)$ into disjoint components. Let $\{\cD_k\}_{k=1}^{M}$ be these duration groups. Continuing the derivation in \eqref{eq:backdoor}, we estimate the deconfounded model $ \bE[W|do(U,V)]$ via the approximation:
\begin{equation}
\begin{split}
      \bE[W|do(U,V)] & =  \sum_{d}\bP(D=d)\bE[W|U,V, D=d]\\
      & \approx \sum_{k=1}^{M}\mathbf{1}\{d\in \cD_k\} \bE[W|U,V, D\in  \cD_k]\\
      &\stackrel{\Delta}{=} \sum_{k=1}^{M}\mathbf{1}\{d\in \cD_k\} f_k(U,V),
\end{split}
\end{equation}
where for each duration group $\cD_k$, $f_k(u,v)$ is the watch-time prediction model fitted on samples $\{(u_i, v_i, w_i, d_i): d_i\in \cD_k\}$ that belong to the duration group.

We hereby provide an intuitive explanation on why such data splitting procedure based on  duration can mitigate the bias issue of  edge $D\rightarrow V$ in Figure \ref{fig:causal}(a). In standard watch-time prediction models such as WLR \cite{covington2016deep}, samples with long watch-time weights more in the gradient updating, such that the prediction model often performs poorly on samples with short watch-time. Watch-time is highly correlated with duration, as shown in Figure \ref{fig:duration-watchtime}. By splitting data based on duration and fitting model group-wisely, we alleviate the interference from  samples with long watch time  on  samples with short watch time during model training.

However, such data-splitting approach raises another concern. If we fit an individual watch-time prediction model $f_k$ for each duration group $\cD_k$ (as demonstrated in Figure \ref{fig:model1}(a)), model size will grow undesirably large, which is not practical in real production systems for scalability concern. But if we allow parameter sharing across duration groups, fitting with the original watch-time labels is equivalent to  learning  without data splitting, which fails to bring in the benefits of duration deconfounding. 
The following section explains how to address this dilemma via transforming the original watch-time labels into duration-dependent watch-time labels, allowing us to both remove  duration bias and also maintain a single set of model parameters to gain scalability.

\subsection{Estimating Watch-time per Duration Group}
\label{sec:estimate}
Moving on, we describe how we fit a single watch-time prediction model using data from all duration groups. Recall that the goal of our design is twofold: (i) duration debiasing and (ii) parameter sharing. The key is to transform the watch-time label to be duration-dependent, manifested by fitting watch-time quantiles---instead of the original values---with respect to the corresponding duration group. Collectively, we introduce  our \textbf{D}uration-\textbf{D}econfounded \textbf{Q}uantile-based (D2Q) watch time prediction framework.

 Denote $\widehat{\Phi}_k(w)$ as the empirical cumulative distribution of watch time on videos in duration group $\cD_k$. Given a user-video pair $(u, v)$, the D2Q method predicts its watch-time quantile in the corresponding duration group and then maps it to the value domain of watch time using $\widehat{\Phi}_k$. That is,
\begin{equation}
    f_k(u,v) = \widehat{\Phi}_k^{-1}(h(u,v)), 
\end{equation}
where $h$ is a watch-time quantile prediction model fitted on data across \emph{all} duration groups:
\begin{equation}
    h = \arg\min_{h'}  \sum_{\{(u_i, v_i, w_i)\}_{i=1}^n}\big(h'(u_i, v_i) - \widehat{\Phi}_{k_i}(w_i)\big)^2,
\end{equation}
with $k_i$ being the duration group of sample $i$ such that $d_i \in \cD_{k_i}$.  One can apply any off-the-shelf regression model to fit the quantile prediction model $h$ and maintain one single set of model parameters across all duration groups. Then, during the inference phase, when a new user-video pair $(u_0,v_0)$ arrives, the model firstly finds the duration group $\cD_{k_0}$ that video $v_0$ belongs to, and then maps the watch-time quantile prediction $h(u_0, v_0)$ to the watch-time value $\widehat{\Phi}_{k_0}^{-1}(h(u_0, v_0))$.  We summarize the learning and inference procedures in Algorithm \ref{alg:training} and \ref{alg:inference} respectively. 

\begin{figure*}
    \centering
    \includegraphics[width=0.95\textwidth]{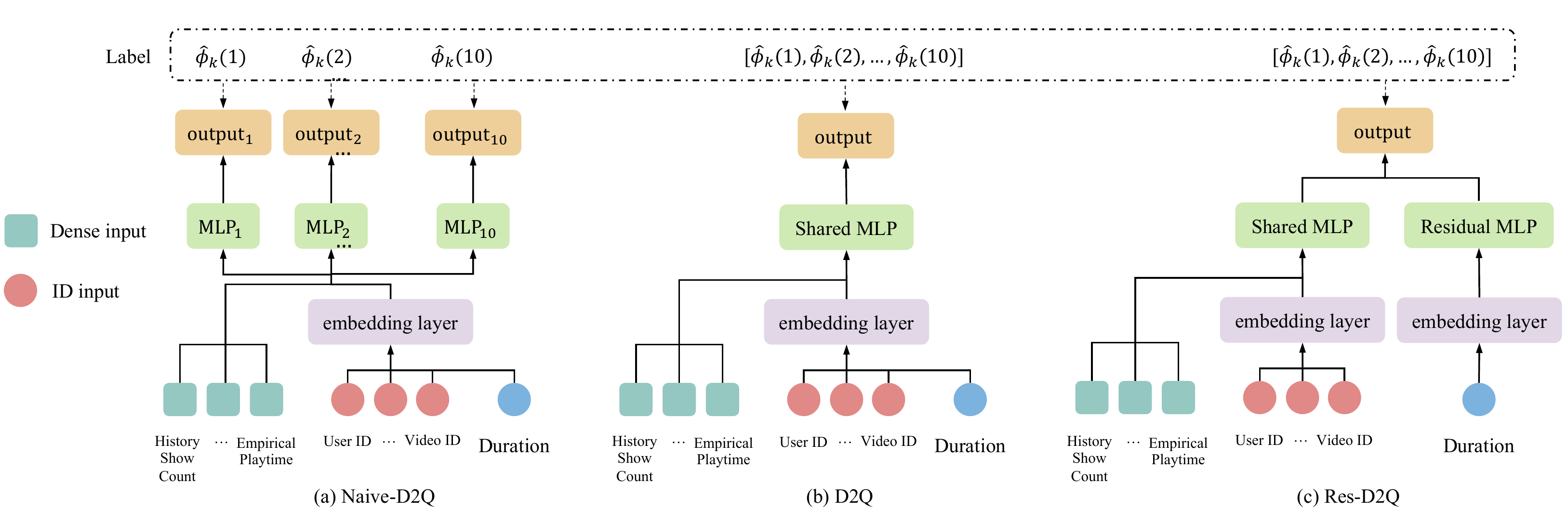}
    \caption{Different model architectures for estimating watch-time per duration group, \textit{i.e.,} $\widehat{\Phi}_k(h(u,v))$. Figure (a)  fits individual models to estimate watch-time per duration group. ``Dense input'' refers to historical statistical numbers (such as \textit{historical show count, empirical watchtime}) of the video. ``ID input'' refers to ID features (such as \textit{user id, video id}) and categorical features (such as \textit{video category, user gender}).  Figure (b) fits a single model across all duration groups, with labels of watch-time quantiles calculated via watch-time empirical distribution in the corresponding duration group. Figure (c) further utilizes duration information in the network architecture and consequently improves watch-time estimation. }
    \label{fig:model1}
\end{figure*}

In this way, D2Q fit labels that are duration-dependent. We note that video duration should also be part of model  input to differentiate samples from different duration groups, as illustrated by Figure \ref{fig:model1}(b). Otherwise, samples from different duration groups may share the same label of watch-time quantile but have different characteristics---a single  model  would fail to learn the watch-time quantile across  groups. To fully utilize duration information, one can additional incorporate a duration adjustment tower in the model architecture as ResNet \cite{he2016deep}, for which we refer to as Res-D2Q and illustrate in Figure \ref{fig:model1}(c). Section \ref{sec:experiment} demonstrates that Res-D2Q further improves watch-time prediction accuracy  over D2Q.

The transformation of watch-time labels with respect to duration  allows for both deconfounding duration bias and parameter sharing across duration groups. However, with the number of duration groups increasing, the group sample size shrinks, and the empirical cumulative distribution of watch-time per duration group will gradually deviate from its true distribution. Therefore, the model performance should firstly improve with duration-based data splitting, thanks to the benefits of deconfounding duration; then with the number of duration groups growing, the estimation error of empirical watch-time distribution will dominate the model performance and make it progressively worse. Section \ref{sec:experiment} empirically justifies this performance change with a series of experiments.

\section{Experimental Results}
\label{sec:experiment}

In this section, we provide empirical evidence to demonstrate the effectiveness of our approach on both real-world data and live experiments. Extensive offline evaluation shows that our method outperforms existing baselines by providing more accurate watch-time prediction, such that the ranking order induced by the predicted values is closer to the ideal ranking. We note that with  the platform's goal of improving user watch time,  ranking---in contrast to the true watch-time value---is usually much valued when recommending videos in practice.   Furthermore, by incorporating our method in the recommender system of a short video platform, we find it effectively  improves real-time video consumption as compared to alternatives, thanks to its ability to generate better ranking of candidate videos based on optimized watch-time prediction. 

\subsection{Offline Evaluation}
\label{sec:offline}
We first evaluate our approach as well as other baselines on offline data collected from real applications~\footnote{Reproduction code can be found at \url{https://github.com/MorganSQ/Ks-D2Q}}. In particular, we are interested in understanding: (i) \emph{how does deconfounding duration contribute to watch-time prediction?}; and (ii) \emph{how does the number of duration groups affect our model performances?}

\subsubsection{Data}
We use production data collected from online recommender systems on the Kuaishou App. By the nature of full-screen feed recommendation, every video in the collected sample has been shown to a user and is associated with the user watch time (which can be close to zero if the user immediately scrolls down to the next video). Specifically, for the causal graph shown in Figure \ref{fig:causal}, we have
\begin{itemize}
    \item user representation $U$: user instantaneous context (such as location, time, and device), stationary context (such as demographics if available), and historical interactions that encode his/her interests.
    \item video representation $V$: video topic information, its corresponding video-creator information, and its previous interactions with other users.
    \item duration $D$: length of the video.
    \item watch-time $W$: the  watch time from the user. 
\end{itemize}
All algorithms evaluated share the same input features.
In total, we have 1,211,885,691 samples for training and 134,653,965 samples for testing, with statistics summarized in Table~\ref{tab:dataset}.

\begin{table}[t]
    \centering
    \caption{Summary statistics of Kuaishou dataset used in offline evaluation in Section \ref{sec:offline}}
    \begin{adjustbox}{max width=1.0\linewidth}
    \begin{tabular}{cccc}
    \toprule
          Users & Photos &  Categories & Instances \\
        \midrule
         47,298,353 & 15,187,170 & 5,942 &  1,346,539,657 \\
        \bottomrule
    \end{tabular}
    \end{adjustbox}
    \label{tab:dataset}
\end{table}

\subsubsection{Methods}\label{sec:methods} We focus on the following methods.
\begin{itemize}
    \item \textbf{VR (Value Regression).} This approach predicts the watch-time value directly, via minimizing   mean squared error loss between the predicted value and the actual watch time.
    \item \textbf{WLR (Weighted Logistic Regression) \cite{covington2016deep}.} This approach fits a weighted logistic regression model and uses the learned odds as the predicted watch time. Since there are no unimpressed videos in our case, we determine the binary labels based on whether the watch time surpasses the $q_{60}$-quantile of the empirical watch-time distribution, which is computed on all training samples. Following \cite{covington2016deep}, positive samples are weighted by watch-time and negative samples receive unit weight.   Appendix \ref{appendix:vtr} details this method. 
    \item \textbf{D2Q (Ours).} As described in Section \ref{sec:estimate}, this approach (i) splits data based on duration; and (ii) fits a regression model---with its architecture shown in Figure \ref{fig:model1}(b)---to estimate watch time quantile via  mean squared error loss. Then the predicted quantile is  mapped to the watch-time value domain---based on the group-wise empirical watch-time distribution---to output the final  watch-time estimation. 
    \item \textbf{Res-D2Q (Ours).} This method further utilizes the duration information by improving D2Q and incorporating duration in model network layers, following the design of ResNet. The model architecture is illustrated in Figure \ref{fig:model1}(c).
\end{itemize}
All algorithms share the same network architecture, except that for the classification-based algorithm WLR and quantile-prediction algorithms D2Q and Res-D2Q, we rescale the output  via a sigmoid function to be within $[0,1]$; and for Res-D2Q, we add a residual multi-layer perception (MLP) for duration adjustment in the last layer to help the model differentiate samples from different duration groups. Appendix \ref{appendix:experiments} specifies the detail of network architecture.  For both of our duration-deconfounded algorithms D2Q and Res-D2Q, we vary the number of duration groups across $[1, 10, 20, 30, 50, 100]$ to study its influence on model performance.

\subsubsection{Metrics} We consider the following metrics for performances.
\begin{itemize}
    \item \textbf{MAE (Mean Absolute Error)}, which measures the mean absolute error between the predicted and true values,
        \begin{equation}
    MAE = \frac{1}{n} \sum_{i=1}^{n}{|y_i - \hat{y}_i}|,
    \label{eq:mae}
\end{equation}
    where $y_i, \hat{y}_i$ are the true and predicted  values for sample $i$.
    \item \textbf{XAUC}, which is an extension of AUC to dense values. For a pair of samples, we score $1$ if the predicted watch-time values of the two videos are in the same order as the ground truth and score $0$ vice versa.  We uniformly sample such pairs from test set and average those scores as XAUC. Intuitively, XAUC measures how the ranking induced by the predicted watch time is in agreement with the ideal ranking. Larger XAUC suggests better model performance.
    \item \textbf{XGAUC}, which calculates XAUC per user and then averages scores with weights proportional to user sample size. Larger XGAUC suggests better model performance.
\end{itemize}
We note that the  ranking-order-related metrics such as  XAUC and XGAUC are often valued more in real applications than the absolute value accuracy measured by MAE, since platforms generate recommendations based on the ranking of predicted values.

\begin{table}
\caption{Offline evaluation results on Kuaishou dataset.  
We use bold fonts to label best performances.}
\centering
\begin{tabular}{lcccc}
\hline
\multirow{2}{*}{\#Groups} & \multirow{2}{*}{Method}  & \multicolumn{3}{c}{Kuaishou dataset.} \\
\cline{3-5}
  & &  XAUC & XGAUC &  MAE \\ 
\hline
\multirow{3}{*}{1}
& VR & 0.6843 & 0.6380 & 28.2413  \\
& WLR\cite{covington2016deep}   & 0.6940  & 0.6436 & 65.6535  \\
& D2Q & 0.7084 & 0.6562 & 36.4871 \\
\hline
 \multirow{2}{*}{10} & D2Q   & 0.7092 & 0.6579 & 26.8684 \\
 & Res-D2Q & 0.7108 & 0.6604 & 26.6686 \\
\hline
 \multirow{2}{*}{20} & D2Q   & 0.7147 & 0.6654 & 26.5993 \\
 & Res-D2Q & 0.7150 & 0.6679 & 26.5530 \\
\hline
 \multirow{2}{*}{30} & D2Q   & 0.7148 & 0.6660 & 26.5227 \\
 & Res-D2Q & \textbf{0.7145} & \textbf{0.6693} & \textbf{26.5073} \\
\hline
 \multirow{2}{*}{50} & D2Q   & 0.7142 & 0.6619 & 26.8047 \\
 & Res-D2Q & 0.7141 & 0.6643 & 27.1260 \\
\hline
 \multirow{2}{*}{100} & D2Q   & 0.7119 & 0.6608 & 26.8277 \\
 & Res-D2Q & 0.7125 & 0.6637 & 26.7297 \\
\hline
\end{tabular}
\label{tab:offline-exp}
\end{table}

\begin{figure}
\centering
\resizebox{\linewidth}{.8\linewidth}{
\begin{tikzpicture}
\begin{axis}[
    xlabel={\# Duration groups},
    ylabel={XGAUC(\%)},
    xmin=1, xmax=6,
    ymin=63, ymax=68,
    xtick={1,2,3,4,5,6},
    xticklabels={1,10,20,30,50,100},
    ytick={63,64,65,66,67,68},
    legend pos=north west,
    ymajorgrids=false,
    grid style=dashed,
]
\addplot[
    color=blue,
    mark=square,
    ]
    coordinates {
    (1,65.62)(2,65.79)(3,66.54)(4,66.60)(5,66.19)(6,66.08)
    };

\addplot[
    color=orange,
    mark=halfcircle,
    ]
    coordinates {
    (1,65.62)(2,66.04)(3,66.79)(4,66.93)(5,66.43)(6,66.37)
    };
\addplot[
    color=black,
    dashed
    ]
    coordinates {
    (1,64.36)(6,64.36)
    };
\addplot[
    color=black,
    dashed
    ]
    coordinates {
    (1,63.80)(6,63.80)
    };    
\node[name=e1] at (axis cs:3.4,64.66) {WLR (64.36$\%$)};
\node[name=e2] at (axis cs:3.4,63.50) {VR (63.80$\%$)};
\draw[->] (e1.east) -- +(3pt,-10pt);
\draw[->] (e2.east) -- +(3pt,10pt);

\legend{D2Q,Res-D2Q}    
\end{axis}
\end{tikzpicture}
}
\caption{XGAUC of watch-time prediction methods with respect to number of duration groups. Our models D2Q and Res-D2Q significantly outperform the baseline methods WLR and VR. Our model performance first improves gradually  with increased number of duration groups,  benefiting from deconfounding duration. Then the performance drops when the group number is too large, due to reduced group-wise sample size and accumulated estimation error. }
\label{fig:offline-eval}
\end{figure}
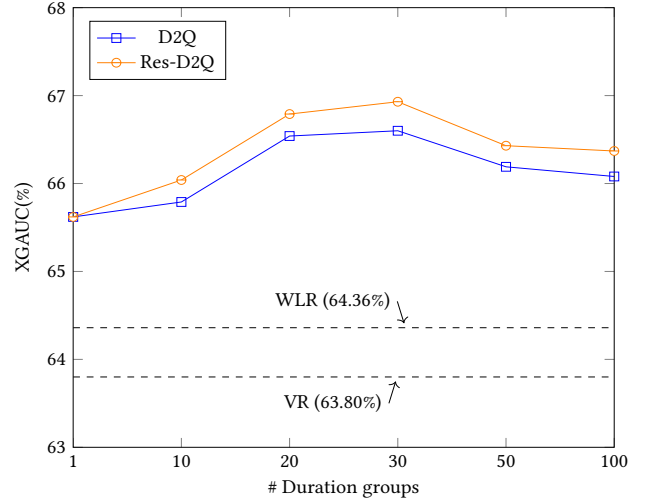

\begin{table*}[t]
\caption{Live experiments on Kuaishou App. 
We use VR as a baseline and show the relative performance  of WLR and Res-D2Q with $\#Groups=30$.  
The square brackets represent the $95\%$ confidence intervals for online metrics. 
Statistically-significant improvement (whose value is not in the confidence interval) is marked with bold font in the table. 
}
\centering
\begin{tabular}{lccccc}
\toprule
 \multirow{2}{*}{Method}  & Main Metric. &  \multicolumn{4}{c}{Constraint Metrics.} \\
\cmidrule(lr){2-2}
\cmidrule(lr){3-6}
  & Watch Time &  Like & Follow & Share & Comment \\ 
\midrule
\multirow{2}{*}{WLR \textit{v.s.} VR (baseline)} & \textbf{+0.184$\%$} & \textbf{+1.012$\%$} & +0.214$\%$ & +0.959$\%$ & -0.137$\%$  \\
& \color{gray}{$[ -0.16\% , 0.16\% ]$} & \color{gray}{$[ -0.50\% , 0.51\% ]$}  & \color{gray}{$[ -0.4\% , 0.4\% ]$} & \color{gray}{$[ -1.31\% , 1.40\% ]$} & \color{gray}{$[ -0.75\% , 0.73\% ]$} \\
\multirow{2}{*}{Res-D2Q \textit{v.s.} VR (baseline)}  & \textbf{+0.746$\%$}  & +0.251$\%$ & -0.167$\%$  & -0.861$\%$  & +0.271$\%$  \\
& \color{gray}{$[ -0.15\% , 0.15\% ]$} & \color{gray}{$[ -0.41\% , 0.41\% ]$}  & \color{gray}{$[ -0.6\% , 0.6\% ]$} & \color{gray}{$[ -1.21\% , 1.21\% ]$} & \color{gray}{$[ -0.85\% , 0.86\% ]$} \\
\bottomrule

\end{tabular}
\label{tab:online-exp}
\end{table*}

\subsubsection{Result-I: Overall performance} Table \ref{tab:offline-exp} shows the performances of different methods with respect to various number of duration groups. Note that there is no data splitting for VR and WLR, and thus we present their results in the row of group number equaling one; when there is only one group, Res-D2Q is equivalent to D2Q since all samples share the same duration adjustment, and thus we omit the result of Res-D2Q there.

Our approaches D2Q and Res-D2Q with $30$ duration groups peak the performance in all metrics XAUC, XGAUC, and MAE. In particular, by further leveraging the duration information in the model architecture, Res-D2Q can better distinguish samples from different duration groups and thus outperforms D2Q in most cases. 
When there is no data splitting, D2Q (which fits the watch time quantile directly) has  comparable performance with LR (which fits the watch time value). However, once data is split based on duration, D2Q with any experimented number of duration groups generates more accurate prediction than LR, endorsing the effectiveness of our  duration-wise data splitting to deconfound duration.

\subsubsection{Result-II: Effect of number of duration groups}
Figure \ref{fig:offline-eval} plots the XGAUC of our methods D2Q and Res-D2Q with respect to number of duration groups. When there is no data splitting, both methods are equivalent to each other. Once data is split to deconfound duration, Res-D2Q is superior to D2Q via improved network architecture with duration information. With the number of duration groups increasing, the performance firstly improves, with the merit of deconfounding duration by data splitting, and then declines, due to the increased estimation error of empirical watch-time distribution resulted from the  shrinkage of sample size. Such observation is in agreement with  the discussion in Section \ref{sec:estimate}.

\subsection{Live Experiments}
\label{sec:live}
We further test our approach in live A/B experiments powered by Kuaishou video recommendation platform, demonstrating its advantage over alternatives in improving real-time video consumption, as a consequence of improved watch-time prediction. 

\subsubsection{Compared methods}
We compare baseline methods VR and WLR with our approach Res-D2Q with $30$ duration groups, which  achieves the best performance in offline evaluation in Section \ref{sec:offline}. Detailed description of methods can be found in Section \ref{sec:methods}. We use the same model architecture as that in Section \ref{sec:offline}.

\subsubsection{Experimental setup} We integrate watch-time prediction into the ranking phase of the online recommender system used by  Kuaishou App. Specifically, when a user arrives, the recommender system first generates a set of candidate videos that the user might be interested in, based on his/her characteristics. Then, the prediction model predicts the watch time of each video candidate supposing that it were recommended to the user. The candidate videos are  ranked in accordance with the predicted values, and videos with higher values have larger likelihood to be recommended.

Models are pretrained on shared samples collected within one day before being tested in real time. We conduct A/B experiments to  evaluate their live performances---we randomly split users on the platform into buckets  of $5\%, 5\%, 5\%, 85\%$ and use the first three buckets for online evaluation, \emph{i.e.}, $5\%$ of the users on Kuaishou App experience the recommender system that embodies the corresponding watch-time prediction model. Considering that Kuaishou serves over $320$ million daily active users \cite{kuaishoureport}, doing experiments in the $5\%$-bucket affects  a huge population of more than $15$ million users, which endorses the statistical significance of our results.

\subsubsection{Metrics} We evaluate model performance based on the total amount of time spent by users in the bucket  on watching the videos, denoted as \watchtime. 
This metric is positively correlated with  watch-time prediction accuracy. Consider a  mental thought:
there are two watch-time prediction methods model-A and model-B, and suppose that model-A better predicts watch time. Then videos with larger ground-truth watch time are more likely to be ranked higher and  consequently shown to users by the recommender system with model-A, as compared to that with model-B;
as a result, users in the bucket that tests model-A will spend more time on watching the videos, improving the \watchtime~metric. 

We also provide additional metrics  widely-adopted in real production systems, which measure user engagement from the  perspective of user interactions that are also valued by platforms, including \emph{like} (clicking like button on the screen), \emph{follow} (following the corresponding video-creator), \emph{share} (sharing the video with friends), and \emph{comment} (leaving comments on the video). However, since interactions are not mutually exclusive among platforms (contrary to \watchtime), platforms usually use them as constraint metrics when evaluating strategies that improve \watchtime. 
We present the total number of like/follow/share/comment counts for each method to complement our evaluation.

\subsubsection{Results}
Table \ref{tab:online-exp} shows the performance of evaluated methods after being tested online concurrently for 24 hours on Kuaishou App. We use VR as a base method and list the relative performances of WLR and Res-D2Q. For the main metric \watchtime, both WLR and Res-D2Q outperform VR with statistical significance, and our method Res-D2Q achieves a larger improvement of $+0.746\%$, which is remarkable given the fact that the average watch-time improvement from production algorithms is around $0.1\%$. Indeed, our approach has been deployed to the online recommender system on Kuaishou App. For the constraint metrics measuring user interactions, the difference between Res-D2Q and VR is not statistically significant and thus can be safely ignored.

\section{Conclusion}
The surge in video consumption has been revolutionizing social media worldwide, causing increasing demand for optimizing the recommender systems on these video platforms. 
It remains to be a key problem to accurately predict the watch-time when evaluating different candidate videos, such that videos with potentially larger watch time are recommended to improve user engagement. 
However, watch-time prediction has been plagued by duration bias that is overlooked in existing models.  Such bias originates from the goal of recommender systems to improve user watch time, such that spurious correlation between duration and watch time may be over-utilized and videos with long duration would be unfairly favored despite that they may be less aligned with user interests. Indeed, many video-sharing platforms are progressively  biased towards videos with long duration. This issue becomes more severe due to the feedback loop of recommender systems, resulting in bias amplification that harms personalization. 

In this paper, we formulate the problem of watch-time prediction using a causal graph, which characterizes the confounding effects of duration on both video exposure and watch-time prediction. We propose a \emph{D}uration-\emph{D}econfounded \emph{Q}uantile-based (D2Q) framework, such that the natural effect of duration on watch-time is preserved, and the bias of duration on video is removed. Through extensive offline evaluation and live experiments, we demonstrate the advantages of our approach over alternatives in providing more accurate watch-time estimations, which further improves video consumption in real-time recommendation on Kuaishou App. We also vary the number of duration groups and show that our model performance improves first and then declines, due to deconfounding duration and reduced sample size respectively. 



\bibliographystyle{ACM-Reference-Format}
\bibliography{paper}

\appendix
\section{Weighted Logistic Regression for Watch-time Prediction}
\label{appendix:vtr}
\subsection{YouTube Method}
Watch-time prediction via weighted logistic regression is first proposed by YouTube~\cite{covington2016deep}, where the expected watch time  is calculated via the intermediate metric \textit{Odds}. For completeness, we review their method here. Odds is defined by Equation~\ref{eq:odds}, where $k$ is the number of ``positive'' (which shall be defined shortly) samples  and $n$ is total number of samples. $T_i$ is the watch time for ``positive'' video $i$. In YouTube recommender system, ``positive'' videos are those that are clicked by users and thus impressed,  and ``negtive'' indicates the recommended video is not clicked. $p_w$ is the predicted watch time.
\begin{equation}
    {\rm Odds} = \frac{\sum_i T_i}{N-k} = \frac{p_w}{1-p_w}.
    \label{eq:odds}
\end{equation}

The calculation of \textit{Odds} can be connected with that of expected watch time $\mathbb{E}(T)$ using Equation~\ref{eq:ctr}. $p_{ctr}$ is the click-through-rate of the YouTube recommender system.  

\begin{equation}
\begin{split}
    {\rm Odds} &= \frac{\sum_i T_i}{N-k} =\frac{\sum_i T_i/N}{(N-k)/N} =\frac{\mathbb{E}(T)}{1-k/N}\\
    &=\frac{\mathbb{E}(T)}{1-p_{\rm ctr}}.
\end{split}
\label{eq:ctr}
\end{equation}

The Taylor expansion~\cite{taylor} of $\frac{1}{1-x}$ is $1+x+x^2+x^3+\cdot\cdot\cdot$. Thus Equation~\ref{eq:ctr} can be rewritten as  Equation~\ref{eq:appro_ctr}, where the approximation is satisfied because $p_{ctr}$ is around $0.01$ such that $p_{ctr}^2$ is small enough.

\begin{equation}
\centering
\begin{split}
    {\rm Odds} &= \frac{\mathbb{E}(T)}{1-p_{\rm ctr}} \\
    &= \mathbb{E}(T) * ( 1 + p_{\rm ctr} + p_{\rm ctr}^2 + p_{\rm ctr}^3 + \cdot\cdot\cdot)\\
    &\approx \mathbb{E}(T) * ( 1 + p_{\rm ctr})\\
    &\approx \mathbb{E}(T).
\end{split}
\label{eq:appro_ctr}
\end{equation}

Combining Equation~\ref{eq:appro_ctr} and Equation~\ref{eq:odds}, the expected watch time can be calculated with Equation~\ref{eq:final_odds}.

\begin{equation}
    \mathbb{E}(T) \approx \frac{p_w}{1-p_w}.
    \label{eq:final_odds}
\end{equation}

\subsection{WLR: Adapting YouTube Method to Our Scenario}
In contrast to the YouTube  scenario where users have ``click'' action and then ``watch''~\cite{covington2016deep},  short video platforms such as Kuaishou App serve users in a ``top-down'' scenario where they  present videos in a full-screen  and single-column format; thus the click-through-rate $p_{ctr}$ in Equation~\ref{eq:ctr} is not well-defined. Instead, we determine the ``positive'' and ``negtive'' labels  based on whether the watch time surpasses the $q_{60}$-quantile of the empirical watch-time distribution. Following \cite{covington2016deep}, positive samples are weighted by watch-time and negative samples receive unit weight. We can get similar result as Equation~\ref{eq:ctr}, which is shown in Equation~\ref{eq:pwtd}.

\begin{equation}
\begin{split}
    {\rm Odds} &= \frac{\sum_i T_i}{N-k} =\frac{\sum_i T_i/N}{(N-k)/N} =\frac{\mathbb{E}(T)}{1-k/N}\\
    &=\frac{\mathbb{E}(T)}{1-p_{\geq q_{60}}}.
\end{split}
\label{eq:pwtd}
\end{equation}

Unfortunately, the average $p_{\geq q_{60}}$ is 0.4 and can not be approximated as $p_{ctr}$ like Equation~\ref{eq:appro_ctr}. Thus the final expected watch time in ``top-down'' scenario can be obtained by Equation~\ref{eq:pwtd}. 
\begin{equation}
    \mathbb{E}(T) = \frac{p_w}{1-p_w}(1-p_{\geq q_{60}}),
    \label{eq:pwtd}
\end{equation}
where $p_{\geq q_{60}}$ is the predicted probability of the watch time surpassing the $q_{60}$ quantile and can be obtained via minimizing the classical log-loss (see Equation~\ref{eq:logloss}) for binary classification.

\begin{equation}
\begin{split}
    \mathcal{L}(\theta) &= - \frac{1}{N} \sum_{j=1}^{N}\Big(~ y_{\geq q_{60}}(j) *log(p_{\geq q_{60}}(j)) \\
    &+ (1-y_{\geq q_{60} }(j))*log(1-p_{\geq q_{60}}(j)) \Big).
\end{split}
\label{eq:logloss}
\end{equation}

\section{Network Architecture in Experiments}
\label{appendix:experiments}
Recall that all methods in our experiment share the same network structure except for Res-D2Q that additionally has a MLP for duration adjustment. 
The overall network architecture can be divided into three parts: input layer, encoding layer, and output layer. 

\subsection{Input Layer}
The goal of input layer is to embed different features, which can be divided into two categories: dense input $x_{dense}$ and id input $x_{id}$. Particularly, we use  $x_{duration}$ to denote the raw value of video duration for its importance in our work.
The output of  the embedding layer is $\bm{E}_{dense} \in  \mathbb{R}^{B \times 32}$, $\bm{E}_{id} \in  \mathbb{R}^{B \times 512}$ and $\bm{E}_{duration} \in  \mathbb{R}^{B \times 32}$, where $B$ is the size of mini batch. In our experiments, we set $B=512$.

\subsection{Encoding Layer and Output Layer}
The output of input layer is concatenated together, which is further transformed by a projection matrix $\bm{W} \in \mathbb{R}^{576 \times 512}$ into an embedding matrix $E$:
\begin{equation}
    \bm{E} = \text{Concat}(\bm{E}_{dense},\bm{E}_{id},\bm{E}_{duration})\bm{W}.
\end{equation}

Then the embedding matrix $\bm{E}$ is fed into three-layer of MLP whose embedding size is $\{256,128,64\}$ respectively, generating a hidden matrix $\bm{H} \in  \mathbb{R}^{B \times 64}$:
\begin{equation}
    \bm{H} = \text{Swish}\Big(\text{Swish}\big(\bm{E}\bm{W}^1 + \bm{b}^1\big)\bm{W}^2 + \bm{b}^2\Big)\bm{W}^3 + \bm{b}^3.
\end{equation}

Finally, $H$ is fed into the output layer, which is output directly for VR and is activated with sigmoid function for others. 
\begin{equation}
    \bm{y} = [y_w(1), y_w(2), \cdot\cdot\cdot]^\top = \text{Sigmoid}(\bm{H}).
\end{equation}
We note that the duration adjustment for Res-D2Q also happens in this layer, where $H$ is additionally concatenated with the output from duration tower.

\end{document}